\documentclass[aps,prd,twocolumn,superscriptaddress,showpacs,floatfix]{revtex4-2}%

\usepackage{graphicx}
\usepackage{tabularx}
\usepackage{bm}
\usepackage{latexsym}
\usepackage{epsf}
\usepackage{rotating}
\usepackage{epsfig,graphics,rotate,color}
\usepackage{wrapfig}
\usepackage{amssymb}
\usepackage{stmaryrd}
\usepackage{amsmath}
\usepackage{amsfonts}
\usepackage{subfigure}
\usepackage{array,hhline,dcolumn}%
\usepackage[normalem]{ulem}
\usepackage{color}
\usepackage{wasysym}
\usepackage[colorlinks=true,allcolors=blue]{hyperref}

\begin{document}
\bibliographystyle{apsrev4-1}

\title{Collapse of Neutrino Wave Functions under Penrose Gravitational Reduction}

\author{B.J.P. Jones}
\affiliation{
University of Texas at Arlington, Arlington, Texas 76019, USA
}

\author{O.H. Seidel}
\affiliation{
University of Texas at Arlington, Arlington, Texas 76019, USA
}

\begin{abstract}
Models of spontaneous wave function collapse have been postulated to address the measurement problem in quantum
mechanics. Their primary function is to convert coherent quantum superpositions into incoherent ones,  with the result that macroscopic objects cannot be placed into widely separated superpositions for observably prolonged times.  Many of these processes will also lead to loss of coherence in neutrino oscillations,
producing observable signatures in the flavor profile of neutrinos at long travel distances. The majority of studies of neutrino
oscillation coherence to date have focused on variants of the continuous
state localization model, whereby an effective decoherence strength
parameter is used to model the rate of coherence loss
with an assumed energy dependence. Another class of 
collapse models that have been proposed posit connections
to the configuration of gravitational field accompanying the mass distribution associated with each wave function that is in the superposition. A particularly interesting and prescriptive model is
Penrose's description of gravitational collapse which proposes a decoherence time $\tau$ determined through $E_{g}\tau\sim\hbar$, where
$E_{g}$ is a calculable function of the Newtonian gravitational potential.
Here we explore application of the Penrose collapse model to neutrino oscillations, reinterpreting previous experimental limits on neutrino decoherence in terms of this model.  We identify effects associated with both spatial collapse and momentum diffusion, finding that the latter is ruled out in data from the IceCube South Pole Neutrino Observatory so long as the neutrino wave packet width at production is $\sigma_{\nu,x}\leq2\times10^{-12}$~m.


\end{abstract}
\maketitle

\section{Introduction}

Neutrino oscillations~\cite{tanabashi2018review} are a consequence of massive and mixed neutrinos
acquiring different quantum phases as they travel over long baselines~\cite{giunti2007fundamentals} . Neutrinos thus represent extremely sensitive quantum interferometers with which to study the structure of spacetime and the test the laws of quantum physics~\cite{icecube2024search,icecube2018neutrino,crocker2004neutrino,stuttard2020neutrino}. No violation of quantum mechanical unitary time evolution has yet been observed at the single particle level, in any system. Despite this, many have argued that they should be a strict necessity in order to resolve the measurement problem in quantum mechanics~\cite{ghirardi1986unified,bassi2013models,weinberg2015lectures}, ultimately providing a sound rational basis for explaining the emergence of definite outcomes when conscious observers interact with the Universe~\cite{hameroff2014consciousness,hameroff2017consciousness}. Sensitive searches for this elusive ``objective reduction'' or collapse process using fundamental particles are thus highly motivated.

Neutrinos have been used to search for anomalous decoherence in a variety of experiments, and these analyses would in principle be sensitive to such violations~\cite{adler2000remarks,hooper2005probing,Gomes:2020muc, de2023neutrino, lisi2000probing,icecube2024search,carpio2019testing}. The results of these experiments are typically interpreted under continuous state reduction models~\cite{bassi2013models,donadi2013effect}.  In this paper, we study the implications of these negative results for collapse theories based on gravitational mechanisms, specifically the model advanced by Penrose in Ref.~\cite{penrose1996gravity}. While predicting broadly similar phenomena to continuous localization models, the mechanics outlined in Ref.\cite{penrose1996gravity} introduce additional subtleties into interpretation of the results in terms of the underlying model parameters. In particular, the collapse rate becomes dependent on the detailed geometry of the neutrino production process, which differs between experiments with different neutrino sources.  At this time, we claim that enough is known about the expected wave packet sizes in neutrino oscillation experiments~\cite{jones2023width,jones2015dynamical,eCap,daya2017study,smolsky2024direct,arguelles2023impact} that we can confront the gravitational collapse model with data from contemporary neutrino oscillation experiments directly.

This paper is structured as follows. Section~\ref{sec:Decoh} schematically reviews the various possible sources of incoherence and decoherence in neutrino oscillations, briefly reviewing the Penrose prescription for gravitational collapse.  Section~\ref{sec:NeutrinoMechanisms} calculates the effects of the gravitational collapse model that are expected in neutrino oscillations.  All of the relevant effects depend on the initial neutrino wave packet width, and we find upper bounds for the value that would be required for the effect to be observable in each type of experiment considered.  Section~\ref{sec:Effect-in-various} compares the required wave packet sizes for each decoherence source against the theoretically expected values in a variety of neutrino experiments.  Of those considered, only the IceCube South Pole Neutrino Observatory~\cite{icecube2024search} has sensitivity to the Penrose model, via the effects of momentum de-localization associated with position-space collapse. The IceCube data does not support this model, as long as the wave packet size at production is smaller than $\sigma_{\nu,x}\leq 2\times 10^{-12}$~m, which is consistent with expectations based on past caluclations~\cite{jones2015dynamical}.  Finally, Section~\ref{sec:Conclusions} summarizes our conclusions.

\section{Sources of incoherence and decoherence in neutrino oscillations\label{sec:Decoh}}

In this section, we outline the basic forms of neutrino coherence loss that may be expected in standard- and non-standard neutrino oscillations. Working in a two flavor model for illustration, a neutrino state
$|\psi\rangle$ produced in flavor  $|\nu_{\alpha}\rangle$ at time $t=t_0$
is comprised of a quantum superposition of mass state $|\nu_{1}\rangle$
and $|\nu_{2}\rangle$ with masses $m_{1}$ and $m_{2}$ with mixing
matrix $U$, as
\begin{equation}
|\psi(0)\rangle=|\nu_{\alpha}\rangle\equiv\sum_{i}U_{\alpha i}|\nu_{i}\rangle.
\end{equation}
The neutrino state vector evolves in time according to
\begin{equation}
|\psi(0)\rangle\rightarrow|\psi(t)\rangle=e^{-\frac{i}{\hbar}Ht}|\psi(0)\rangle.
\end{equation}
If we make the simplifying assumption that each $\nu_{i}$ is produced in an energy Eigenstate
with energy $E_{i}$, then this becomes
\begin{equation}
|\psi(t)\rangle=\sum_{i}U_{\alpha i}e^{-\frac{i}{\hbar}E_{i}t}|\nu_{i}\rangle.
\end{equation}
Evaluating the probability the neutrino will once again be found in
state $|\nu_{\alpha}\rangle$ after time $t$ amounts to projecting
back onto the original flavor state to find the survival probability.
For a simple two-flavor scenario we find the answer
\begin{equation}
P_{\alpha\rightarrow\alpha}(t)=|\langle\nu_{\alpha}|\psi(t)\rangle|^{2}=1-\sin^{2}2\theta\sin^{2}\left(\frac{1}{2\hbar}\left[E_{1}-E_{2}\right]t\right).
\end{equation}
In the case where both mass states are in the same momentum basis
state and the neutrino is fully relativistic such that $t=L$, we
find $E_{1}-E_{2}\sim\Delta m^{2}L/2E$, resulting in the standard
formula for neutrino oscillations,
\begin{equation}
P_{\alpha\rightarrow\alpha}(L)=1-\sin^{2}2\theta\sin^{2}\left(\frac{\Delta m^{2}L}{4E}\right).
\end{equation}
We have switched to natural units with $\hbar=c=1$ and will use them for the rest of this paper. While we made an equal-momentum assumption for simplicity of notation,
the same formula is also obtained if the neutrinos are not in equal
momentum states but instead in the expected kinematic states produced
from a common two- or three-body decay, as long as the wave packet separation effects discussed below are not significant.

An often un-stated assumption in this derivation is that it requires
coherence to be maintained between the propagating mass states during
their travel~\cite{akhmedov2009paradoxes}. Loss of coherence amounts to an effective collapse of
the wave function, for example,
\begin{equation}
|\psi(t)\rangle=\sum_{i}U_{\alpha i}e^{iE_{i}t}|\nu_{i}\rangle\rightarrow\begin{cases}
\begin{array}{c}
e^{-iE_{1}t}|\nu_{1}\rangle\\
e^{-iE_{2}t}|\nu_{2}\rangle
\end{array} & \begin{array}{c}
P_{1}=\cos^{2}\theta\\
P_{2}=\mathrm{sin}^{2}\theta
\end{array}\end{cases}
\end{equation}
where the two possible outcomes are realized with probability $P_{1}$
and $P_{2}$, respectively. In this scenario the oscillatory behavior
will be lost, and the flavor composition becoming invariant with distance, fixed at
\begin{equation}
P_{\alpha\rightarrow\alpha}(L)=\sum_{i}P_{i}|\langle\nu_{\alpha}|\nu_{i}\rangle|^{2}=1-\frac{1}{2}\sin^{2}2\theta.
\end{equation}
There are multiple mechanisms by which loss of coherence can take
place. One possibility is environmental decoherence, whereby the neutrino
becomes somehow entangled with an external system $|\epsilon\rangle$,
such that
\begin{equation}
|\psi\rangle\rightarrow\sum_{i}U_{\alpha i}|\nu_{i}\rangle\varotimes|\epsilon_{i}\rangle,\quad\quad|\epsilon_{i}\rangle\neq|\epsilon_{j}\rangle.
\end{equation}
Then the final state accompanying $|\nu_{1}\rangle$ is distinct from
the final state accompanying $|\nu_{2}\rangle$, and the oscillation
probability becomes
\begin{equation}
P_{\alpha\rightarrow\alpha}(L)=1-\sin^{2}2\theta\left[\frac{1}{2}+\frac{1}{2}\mathcal{R}\langle\epsilon_{1}|\epsilon_{2}\rangle\cos\left(\frac{\Delta m^{2}L}{2E}\right)\right].\label{eq:EnvDecoh}
\end{equation}
In the case where the environment entangled with $|\nu_{1}\rangle$
is very different to that entangled with $|\nu_{2}\rangle$, then
$\langle\epsilon_{1}|\epsilon_{2}\rangle=0$ destroying all interference. Equation~\ref{eq:EnvDecoh} makes it clear that partial environmental decoherence is also possible for intermediate values of $\mathcal{R}\langle\epsilon_{1}|\epsilon_{2}\rangle$, which is relevant for scenarios where the environment gradually gains information about the mass states rather than resolving them in a single entangling interaction.
Because neutrinos barely interact with their environments, environmental decoherence is not
expected in any currently accessible experimental configuration. It may
become relevant in exotic scenarios where entanglements develop with
new beyond-standard-model background fields~\cite{nieves2019neutrino}.

A second source of decoherence is wave packet separation. Neutrinos
are not born in perfect momentum Eigenstates, but instead are produced
as wave packets~\cite{akhmedov2009paradoxes,giunti1998coherence}, such that
\begin{equation}
|\psi\rangle=\sum_{i}U_{\alpha i}\int d^{3}p\,\psi_{i}(p)|\nu_{i},p\rangle,
\end{equation}
where $|\nu_{i},p\rangle$ is a mass eigenstate of mass $m_{i}$ and
momentum $p$ and $\psi_{i}(p)$ is the wave function this mass state.
In the proceeding sections we will follow the standard practice of assuming the
wave functions can be well approximated by Gaussian functions, such
that
\begin{equation}
\psi_{i}(p)=\frac{1}{\sqrt{2\pi\sigma^{2}}}\exp\left[-\frac{\left(p-p_{0}^{i}\right)^{2}}{4\sigma_{\nu,p}^{2}}\right].
\end{equation}

In this case the position-space wave function width can be related
to the momentum-space one by the lower bound of the Heisenberg uncertainty
principle,
\begin{equation}
\sigma_{\nu,x}\sigma_{\nu,p}=\hbar/2.~\label{eq:HeisUncert}
\end{equation}
We note that caution must be exercised in applying relation~\ref{eq:HeisUncert} to neutrino wave packets, since $\sigma_{\nu,x}$ is the coherent spatial wave function width,  often significantly smaller than the experimenter's uncertainty about the emission position of the neutrino.   This distinction is discussed in some detail in Ref.~\cite{jones2023width}. 

Because the neutrino mass states are produced in a common decay
process and the kinematics in the final state are distinct in each
case, the central momenta $p_{0}^{i}$ accompanying each mass state
are nonequivalent. As the wave packets propagate, they therefore travel with
different group velocities. This means that eventually the  wave functions accompanying each mass state will
separate, and as they do so, oscillation coherence is lost. This phenomenon
has been discussed at length in many past works, including but not limited to Refs.~\cite{akhmedov2009paradoxes,giunti2002neutrino,beuthe2002towards,giunti1998coherence,jones2015dynamical,jones2023width}. The distance
$L_{coh}$ over which coherence becomes lost is related to the initial
neutrino wave packet width $\sigma_{\nu,x}$ via 
\begin{equation}
L_{coh}=2\sqrt{2}\frac{\sigma_{\nu,x}}{\Delta v_{ij}},\label{eq:CohLength}
\end{equation}
where $\Delta v_{ij}$ is the velocity difference between neutrino
mass eigenstates, $\Delta v=\Delta m^{2}/2E^{2}$.  The major challenge
with understanding the expected observable consequences of this coherence
loss is the prediction of the initial state wave function width $\sigma_{\nu,x}$.
This is a quantity that depends on the kinematics of the decay and
the localization of the initial state by interactions with its environment,
and has been calculated for several systems of interest including
meson decay in accelerator neutrino beams~\cite{jones2015dynamical}, beta decay in nuclear
reactors~\cite{jones2015dynamical} and electron capture sources~\cite{eCap}. Although it also depends on the quantity $\sigma_{\nu,x}$, we stress that this standard ``wave packet separation'' effect is a  distinct coherence loss mechanism to the one that is our main focus in this paper.

The decoherence phenomena described thus far occur for neutrinos
obeying ordinary quantum mechanical time evolution. Additional losses
of coherence may then be present, if there are fundamental violations of the unitary time evolution of quantum mechanics. 

The non-standard collapse theories fall into two broad classes:
spontaneous collapse models such as Ref.~\cite{ghirardi1986unified} which posit that there is a
fundamental law acting to reduce superpositions into incoherent states directly; 
and collapse theories that postulate that the geometry of the gravitational field plays some specific role
in the process. The first class of models is well represented by the
continuous state localization model~\cite{bassi2013models},
under which superpositions are reduced to incoherent sums of their
basis states in some basis at a specified rate. Its impact on neutrino oscillations
has been explored in many works, such as Ref.~\cite{stuttard2020neutrino,de2023neutrino,carpio2018revisiting,gago2002study,barenboim2024quantum}. The effect is a steady
loss of coherence that suppresses off-diagonal elements in the flavor-space
density matrix. Since the precise dynamics of state-localization are
unknown, the effects of spontaneous collapse processes are typically
modeled by introducing a set of operators with power-law energy dependence
into the Lindblad equation for neutrino oscillations. A typical approach is to  set limits on the possible magnitude of the decoherence strength $\Gamma$ that multiplies each relevant Lindblad operator, scaling as 
\begin{equation}
\Gamma=\Gamma_0\times \left(\frac{E}{E_0}\right)^n,\label{eq:GammaScaling}
\end{equation}
with $n$ as an unknown energy exponent~\cite{stuttard2020neutrino} and $E_0$ a pivot energy that is chosen for convenience. The effective dynamics of these
models are often considered as representative of a broad class of
models that include virtual black hole formation in spacetime foam
~\cite{hawking1978spacetime,stuttard2020neutrino}, deformation of symmetries~\cite{Arzano:2022nlo, DEsposito:2023psn}, metric perturbations~\cite{DEsposito:2023psn,Goklu:2009zz},  fluctuating minimal lengths ~\cite{Petruzziello:2020wkd, DEsposito:2023psn} and light cone fluctuations~\cite{stuttard2021neutrino}

Prototypical examples of models invoking the geometry of gravitational field to explain collapse include the Penrose model~\cite{penrose1996gravity},
the Diosi model~\cite{diosi1987universal,bahrami2014role}, and the Karolyhazy~\cite{penrose1986quantum} model. In these approaches,
the distinct spacetime metric curvatures that accompany different
wave function components in superposition lead to a collapse. Penrose
reasons in Ref.~\cite{penrose1996gravity} that the characteristic energy
scale for collapse would be given by 
\begin{eqnarray}
E_{g}&=\frac{1}{G}\int d^{3}x\left(\nabla\psi_{2}-\nabla\psi_{1}\right)^{2},\label{eq:PenroseEg}\\
&=4\pi\int d^{3}r(\psi_{1}-\psi_{2})(\rho_{1}-\rho_{2}),\label{eq:PenroseEq2}
\end{eqnarray}
where $\psi_{i}$ is the gravitational potential associated with each
mass distribution $\rho_i$ in the superposition and $G$ is Newtons
constant. The decoherence time for these two distributions is found
to be 
\begin{equation}
\tau \sim 1/E_{g}.\label{eq:PenroseUncert}
\end{equation}
This is motivated on the basis that states on spacetimes with different metrics will evolve into incompatible Hilbert spaces~\cite{penrose2014gravitization,penrose1996gravity}, and as such, coherently interfering quantum states cannot be supported. Therefore, superpositions of states corresponding to significantly different gravitational curvatures must become collapsed, preventing large objects from existing in prolonged states of macroscopically separated superposition.  

Applying this model to real
materials used in experiments immediately runs into a challenging problem: for mass distributions
that are truly point-like, expression \ref{eq:PenroseEg} diverges.
Penrose~\cite{penrose2014gravitization} and Diosi~\cite{diosi1987universal} propose distinct solutions to this problem. Penrose
suggests that no particle in the real world has a truly point-like wave function, since
evolution of its wave packet will disperse it until the point that
gravity limits this spreading. Thus the prescribed approach is  to
apply the Schrödinger Newton equation~\cite{bahrami2014schrodinger} , a nonlinear form of the Schrödinger equation that incorporates a gravitational self-attraction term between different parts of the wave function. This protocol delivers the equilibrated wave
function width for the particle before collapse, after which Eq. \ref{eq:PenroseEg}
can be applied to the suitably broadened state. Diosi opts for a different solution,
postulating a new fundamental length scale on which the mass distributions
must be smeared in order to apply Eq. \ref{eq:PenroseEg}. The distance scale advocated by Diosi is in principle, an experimentally discoverable and fundamental quantity.

The avoidance of divergences originating from a point-like wave function, however,
is not a relevant concern for any system where particles are produced with a non-trivial
quantum mechanical width. This is the case in neutrino oscillation experiments.
Ref.~\cite{donadi2013effect} suggests that the Penrose model cannot reasonably be applied to neutrino
oscillation system because the solution of the Schrödinger-Newton equation \cite{bahrami2014schrodinger}
leads to meaninglessly short decoherence distances in the case where the neutrino is allowed to reach this fully collapsed width. However, this
misses a crucial point, that even if the neutrino wave packet width would ultimately become limited to the scale set by the Schrödinger Newton equation,
it does not have time to spread to reach this width in experimental
conditions.  Instead, in all terrestrial experiments the neutrino wave packet width remains very close to its initial value during the neutrino flight time, since the dispersion effect is very slow, scaling as $(\Delta m^2)^2/E^4$. As such, for neutrino oscillations we can apply gravitational collapse models without applying an additional smearing effect, since the neutrino source determines the relevant coherent wave packet width. This is distinct from the assumptions made to treat macroscopic objects in Ref.~\cite{penrose2014gravitization}, where it is assumed as a starting point that the equilibrium gravitationally collapsed width has been reached.

Neutrino widths emerging from the production process have now been calculated for many of the scenarios
of experimental interest \cite{jones2023width,jones2015dynamical,eCap}.
These predictions should provide sufficient information to apply the Penrose collapse model to predict neutrino coherence loss distances.  Unlike in the continuous state reduction models, the gravitational collapse model couples the geometrical shape of the neutrino wave packet to its gravitational collapse,  so the effects of standard wave packet separation
and gravitational collapse must be considered together. Both of these effects acting
as a function of $\sigma_{\nu,x}$ with distinct neutrino energy and baseline scalings. In the next section we calculate these effects.

\section{Effects of gravitational collapse in neutrino oscillations~\label{sec:NeutrinoMechanisms}}

The precise details of the collapse dynamics are not specified by Penrose's approach. Nevertheless, the arguments leading
to it can be extended to the neutrino oscillation system with some small and
defensible extrapolations. We first note that for relativistic particles,
gravitational curvature is sourced primarily by the energy rather than
mass, so we consider the densities and potentials in Eq. \ref{eq:PenroseEq2}
to be sourced by the energy-weighted distribution of $|\psi(x)|^2$ of the relativistic neutrino. This approach is consistent with the observation that
in general relativity light rays gravitate towards each other with effective
mass determined by the photon energy \cite{tolman1931gravitational,ratzel2016gravitational}, their energies serving as a source of metric curvature. A highly
relativistic neutrino should behave rather similarly to a photon in terms
of its gravitational dynamics.  We also note that the Penrose prescription is not strictly  Lorentz invariant 
since Eq. \ref{eq:PenroseUncert} involves the product of two timeline
four-vector components.  Since the mechanism purports to address the measurement problem in quantum mechanics, we perform the calculation in the observers rest frame, where whatever is considered to be a ``measurement'' under this framework must presumably be executed.  We now enumerate the effects that the gravitational collapse process is expected to have on coherence of oscillating neutrinos.

First, if we consider that the neutrino wave function is a sum of
different neutrino mass states, each will lead to a distinct spacetime
curvature due to the different magnitude of the mass in each case.
This leads to a difference in self-energy that can be used to estimate
a decoherence time. We call this \textbf{effect~1}, evaluated in Section
\ref{sec:Effect-1:-Collapse}. Second, neutrinos with different masses
travel at different velocities and can separate spatially. This leads
to a second contribution to the decoherence rate expected under the
Penrose model which is more similar to the effects previously explored
by Penrose $et.al.$ for classical systems \cite{penrose1996gravity}
and Bose Einstein Condensates \cite{howl2019exploring}. We call this
\textbf{effect~2}, evaluated in Sec \ref{sec:Effect-2:-Collapse}\textbf{.}
In practice, Effects 1 and 2 should be calculated simultaneously,
and we present this combined calculation in Sec. \ref{sec:Combination-of-Effects}.
There is also a third, somewhat less direct source of coherence loss
expected in this model. Since collapses in the position basis act
to localize the spatial extent of the wave function, Heisenberg's
uncertainty principle demands they must also broaden it in momentum
space. This leads to a stochastic contribution to the oscillation
phase, which we term \textbf{effect~3}, evaluated in Sec.~\ref{sec:Effect-3:-Decoherence}.
In all of these cases, the magnitude of the effect depends on the
spatial extent of the neutrino wave packet at production. Section
\ref{sec:Effect-in-various} evaluates the magnitude of the effect
in various neutrino experiments based on the expected neutrino wave
packet widths therein.

\subsection{Effect~1: Collapse via neutrino mass difference \label{sec:Effect-1:-Collapse}}

A given
initial state produces neutrino mass basis states with slightly different
energies and momenta to one another, with the energy and momentum differences calculable from kinematic considerations. The
energy difference between two mass states produced 
is
\begin{equation}
\Delta E\sim\zeta\frac{\Delta m^{2}}{2E},\label{eq:EnergyDiff}
\end{equation}
where $\zeta$ is an order-1 number that is a function of the masses
of the initial and final states \cite{giunti1998coherence}. To study
\textbf{effect~1} in isolation from the other effects, we consider the case
where the spatial difference between the two neutrino mass state wave
packets is negligible, so $\psi_{1}=\psi_{2}=\psi$. Then we have 
\begin{equation}
E_{g}=\frac{\Delta E_{\nu}^{2}}{G}\int d^{3}x\left(\nabla\psi\right)^{2}.\label{eq:PenroseEg-1}
\end{equation}
The gravitational potential for a mass density $\rho$ is determined
by Poisson's equation,
\begin{equation}
\nabla^{2}\psi=4\pi G\rho.
\end{equation}
If we consider $\rho$ to be a Gaussian distribution of width dictated
by the width of the neutrino wave packet $\sigma_{\nu,x}$ \cite{jones2023width}
then it can be shown that the relevant potential is
\begin{equation}
\Phi=\frac{G}{4\pi}\frac{1}{r}\mathrm{erf}\left(\frac{r}{\sqrt{2}\sigma_{\nu,x}}\right),
\end{equation}
which we can insert into Eq. \ref{eq:PenroseEq2},
\begin{eqnarray}
E_{g}=&\frac{\Delta E_{\nu}^{2}G}{(2\pi)^{3/2}\sigma_{\nu,x}^{3}}\int d^{3}x\frac{1}{r}\mathrm{erf}\left(\frac{r}{\sqrt{2}\sigma_{\nu,x}}\right)\exp\left(\frac{r^{2}}{2\sigma_{\nu,x}^{2}}\right)\\
&=\frac{\Delta E_{\nu}^{2}G}{\sqrt{\pi}\sigma_{\nu,x}}.
\end{eqnarray}
Inserting Eq. \ref{eq:EnergyDiff}, we find that the decoherence time
is, in natural units,
\begin{equation}
\tau\sim 1/E_g=\frac{\sigma_{\nu,x}}{G}\left(\frac{2E_{\nu}}{\Delta m^{2}}\right)^{2}.
\end{equation}
For highly relativistic neutrinos with $\tau\sim L$, this effect
will become significant for wave packet sizes smaller than
\begin{equation}
\sigma_{\nu,x}\leq G\left(\frac{\Delta m^{2}}{2E_{\nu}}\right)^{2}\,L.
\end{equation}
This will be contrasted against expected wave packet sizes in contemporary
experiments in Sec. \ref{sec:Effect-in-various}.

\subsection{Effect~2: Collapse via neutrino spatial separation \label{sec:Effect-2:-Collapse}}

Not only are neutrino mass states produced with slightly different
central energies, but they also travel with slightly different velocities
\cite{akhmedov2009paradoxes}. This leads to a spatial separation
that is a second source of wave function collapse. To estimate the
magnitude of the effect, we can apply the estimate of $E_{g}$ for
hard sphere mass distributions calculated in Ref. \cite{howl2019exploring},
\begin{equation}
E_{g}=\frac{GM^{2}}{R}\left(2\lambda^{2}-\frac{3}{2}\lambda^{3}+\frac{1}{5}\lambda^{5}\right),\quad\quad\lambda=\frac{b}{2R}.\label{eq:bEq}
\end{equation}
where $R$ is the sphere radius, which we take to be the wave
packet width $\sigma_{\nu,x}$ and $b$ is their separation, which
will be given in terms of the velocity difference $\Delta v_{ij}$
by 
\begin{equation}
b=\Delta v_{ij}L=\frac{\Delta m^{2}}{2E^{2}}L.
\end{equation}
Again we will take $M\sim E_{\nu}$, this time assuming a constant
energy for both neutrino packets to investigate only the effect of
their spatial separation an hence isolate \textbf{effect~2}, We thus arrive
at
\begin{equation}
E_{g}=\frac{GE_{\nu}^{2}}{\sigma_{\nu,x}}\left(2\lambda^{2}-\frac{3}{2}\lambda^{3}+\frac{1}{5}\lambda^{5}\right).
\end{equation}
The decoherence time in the Penrose model for small separation $\lambda\ll1$
is given by
\begin{equation}
\tau\sim 1 /E_{g}=\frac{2\sigma_{\nu,x}^{3}}{GE_{\nu}^{2}L^{2}}\left(\frac{2E_{\nu}^{2}}{\Delta m^{2}}\right)^{2}.
\end{equation}
For a relativistic neutrino with $\tau\sim L$ we find the effect
will be large for wave packet sizes that are small relative to their separation,
\begin{equation}
\sigma_{\nu,x}\leq\left[\frac{G}{8E_{\nu}^{2}}\left(\Delta m^{2}\right)^{2}\right]^{1/3}L.
\end{equation}

\subsection{Combination of Effects 1 and 2 \label{sec:Combination-of-Effects}}

To rigorously account for effects 1 and 2, we need to incorporate them both simultaneously. A combined calculation using Gaussian wave packets yields
\begin{eqnarray}
E_{g}&=&E_{g}^{(1)}+E_{g}^{(2)},\\
E_{g}^{(1)}&=&\Delta E^{2}\sqrt{\frac{2}{\pi}}\frac{G}{\sigma_{\nu,x}},\\
E_{g}^{(2)}&=&2E_{\nu}^2G\left(\sqrt{\frac{2}{\pi\sigma^{2}}}-\frac{1}{b}\mathrm{\mathrm{erf}}\frac{b}{\sqrt{2\sigma^{2}}}\right).
\end{eqnarray}
Since we are interested in small values of $b$ we can Taylor expand $E_{g}^{(2)}$ to leading order, which gives
\begin{equation}
E_{g}^{(2)}\sim\frac{1}{3}\sqrt{\frac{2}{\pi}}\frac{E_{\nu}^2 G}{\sigma^{3}}b^{2},\quad b=\left(\frac{\Delta m^{2}}{2E_{\nu}^{2}}\right)L.
\end{equation}
We note that the effective collapse time from $E_{g}^{(1)}$ is identical to our earlier estimate of the contribution from \textbf{effect~1} derived in Sec~\ref{sec:Effect-1:-Collapse}, and the contribution from $E_{g}^{(2)}$ differs from the previous estimate for \textbf{effect~2} from Sec.~\ref{sec:Effect-2:-Collapse} only by a multiplicative numerical factor of $\frac{2\sqrt{2}}{3\sqrt{\pi}}=0.53$, attributed primarily to the use of Gaussian rather than spherical wave packets.  In all cases the magnitude of \textbf{effect~2} encoded in $E_g^{(2)}$ are far larger than magnitude of effect  1 that is encoded in $E_g^{(1)}$. As such, in subsequent sections we will find that the spatial decoherence that arises will always be dominated by effect~2, not by effect~1.

\subsection{Effect~3: Decoherence via momentum delocalization \label{sec:Effect-3:-Decoherence}}

The phase of a neutrino that dictates oscillation is equal to the
difference between the total accumulated phases of each mass state,
which evolve as $e^{-iE_{i}t}$, where $E_{i}=\sqrt{p^{2}+m_{i}^{2}}$.
If the value of $p$ were to change along the journey, the appropriate
expression for the oscillation will instead become
\begin{equation}
\phi_{ij}=\int dt\,\left(E_{i}(t)-E_{j}(t)\right).\label{eq:OscPhaseIntegral}
\end{equation}
A steady localization in the position
basis as generated by the gravitational collapse effect, whatever its microscopic origin, necessarily implies a steady delocalization in the momentum
basis, as required by the Heisenberg principle. As such, if the collapse
process is acting to localize the positions of the neutrino mass states
it must be acting to delocalize their momenta.   While the Penrose model does not provide the specific operators to use to implement this gravitational collapse, the conclusion is derived in a formal and general way in Ref.~\cite{donadi2023collapse}: whenever a Lindblad operator depends on position, the expectation value of momentum is not constant.  As such, the central momentum of the wave function will undergo random fluctuations, if collapses are allowed to occur. We note that such
accelerations are also predicted to give  rise to photon emission when applied to charged
particles, and this effect  has recently been searched for in low background underground experiments, in Refs~\cite{arnquist2022search,donadi2021underground}.

In the neutrino oscillation system, since two mass states are becoming
distinguished by the collapse process, the random momentum perturbations
will be independent for each mass basis state, since if they were not independent,
the result would be the entire wave packet being translated in position
space while maintaining its overall coherence. Thus we must consider
that the central momentum of a given neutrino mass state at the time of measurement is inequivalent
to its energy at the time of emission, such that
\begin{equation}
E_{i}=\sqrt{\left(p_{0}+\delta p_{i}(t)\right)^{2}+m_{i}^{2}}.
\end{equation}
Here we assume $\delta p_{i}$ is a randomly fluctuating function,
with a mean of zero $\langle\delta p_{i}\rangle=0$. Assuming $\delta p_{i},m_{i}\ll E_{\nu}$,
we can expand,
\begin{equation}
E_{i}\sim E_{\nu}+\frac{2E_{\nu}\delta p_{i}+\delta p_{i}^{2}+m_{i}^{2}}{2E_{\nu}}.
\end{equation}
Where $E_{\nu}$ is the mean neutrino energy, averaged over mass basis states.
To find the impact on the oscillation phase, we must evaluate the
integral of Eq. \ref{eq:OscPhaseIntegral},
\begin{equation}
\phi_{ij}=\frac{\Delta m^{2}}{2E_{\nu}}t+\int dt\,\left(\delta p_{i}(t)+\frac{\delta p_{i}^{2}(t)}{2E_{\nu}}\right).
\end{equation}
The second term will tend to average to zero along the journey, because
$\langle\delta p_{i}(t)\rangle=0$. Oscillations will be decohered
when the third term significantly de-phases the standard oscillation
phase. This will be the case whenever
\begin{equation}
\delta\phi\equiv\frac{\int dt\,\langle\delta p_{i}^{2}(t)\rangle}{2E_{\nu}}\sim{\cal O}(1).\label{eq:Anomalousdphi}
\end{equation}
The RMS value of the momentum fluctuations $\Delta p=\sqrt{\langle\delta p_{i}^{2}\rangle}$
is the momentum width that appears in the Heisenberg uncertainty principle due to the collapse process.
To estimate this quantity, we note that the maximal coherent separation
scale on which position superpositions may exist in this model is
given by the $b$ that appears in Eq. \ref{eq:bEq}. As such,
\begin{equation}
\Delta x\leq\sqrt{\frac{2\sigma_{\nu,x}^{3}}{GE_{\nu}^{2}t}}.
\end{equation}
This upper limit on coherent position uncertainty implies a lower
limit to the momentum uncertainty, as
\begin{equation}
\Delta p\geq\frac{1}{2\Delta x}=\frac{1}{2}\sqrt{\frac{GE_\nu ^{2}t}{2\sigma_{\nu,x}^{3}}},
\end{equation}
The anomalous oscillation phase acquired from the dynamical position-space
collapse is, from Eq. \ref{eq:Anomalousdphi},
\begin{equation}
\phi\sim\frac{1}{16}\frac{GE_\nu L^{2}}{\sigma_{\nu,x}^{3}},
\end{equation}
which will be significant whenever
\begin{equation}
\sigma_{\nu,x}\leq\left(\frac{GE_\nu L^{2}}{16}\right)^{1/3},
\end{equation}
fixing the critical scale of $\sigma_{\nu,x}$ from \textbf{effect~3}.

\section{Observability of gravitational collapse in neutrino experiments \label{sec:Effect-in-various}}

All of the effects described above are observable only when the wave packet width at production $\sigma_{\nu,x}$ is suitably small.  Each effect also has a distinct dependence on baseline and energy, motivating consideration of relative  observability in various neutrino experiments.  For the purpose of assessing observability, we  consider the effect to be potentially detectable if it leads to a decoherence distance of at most 10$\times$ the baseline of the experiment, thus yielding a 10\% decohered neutrino flux. This appears to be a conservative criterion for detectability.

\begin{figure}[t]
\begin{centering}
\includegraphics[width=0.99\columnwidth]{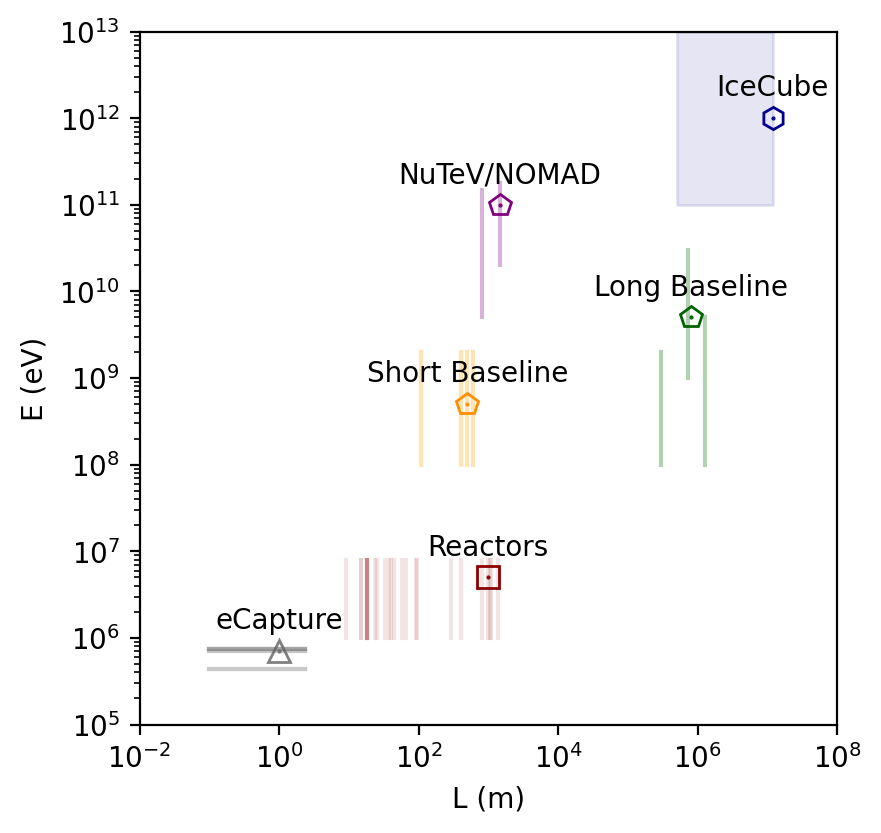}
\par\end{centering}
\caption{Experiments considered in this section. The colored lines and boxes give a rough sketch of the energy and baseline spans of the relevant experiments, and the markers show the values we have taken as representative parameter points for each experiment class for subsequent calculations.\label{fig:Experiments}}
\end{figure}

We compare the following cases, represented in terms of their approximate energy and baseline in Fig.~\ref{fig:Experiments}:

\begin{itemize}
    \item {\bf Electron capture experiments:}  The lowest energy experiments we consider are electron capture experiments.  The BEST experiment~\cite{barinov2022results} uses 50 tons of GaCl$_3$-HCl solution to radio-chemically detect electron neutrinos produced in electron capture decay of $^{51}$Cr.  The experiment observed a defeceit of neutrinos, confirming the previous anomalies of the SAGE and GALLEX experiments~\cite{elliott2023gallium}.  Explanations of the anomaly based on neutrino decoherence have been advanced in Refs.~\cite{krueger2023decoherence,farzan2023decoherence}.  The energies of the neutrinos in BEST emerge at four energies,energies; 747 keV(81.63\%), 427 keV (8.95\%), 752 keV (8.49\%), and 432 keV (0.93\%), shown as grey lines in Fig.\ref{fig:Experiments}. For our estimations we use the flux-weighted mean of these values.  The  experiment is cylindrical, 2.34~m in height and 2.18~m in diameter, and we consider a representative neutrino baseline of 1~m. The expected wave packet width in electron capture decays has been estimated for $^7$Be to be 2.7~nm in Ref.~\cite{eCap}, and a similar method can be used to predict the wave packet width in $^{51}$Cr electron capture to yield a value of $\sigma_{\nu,x}\sim70$~pm. The BeEST collaboration has recently published an experimental lower limit on the neutrino wave packet width in $^7$Be decay of $\sigma_{\nu,x}\geq2.7$~pm, still significantly smaller than the predicted value.
    
    \item {\bf Nuclear reactors}: Nuclear reactors are copious neutrino sources.  A wide variety of nuclear reactor neutrino experiments have operated at many different baselines~\cite{wen2017reactor}. An indication of non-standard reactor anti-neutrino disappearance was observed by comparing  anti-neutrino fluxes from reactors to calculations, and finding the data to be anomalously low~\cite{mention2011reactor}. However, this anomaly has been confronted by new calculations of reactor fluxes that ameliorate the issue significantly~\cite{zhang2024reactor}. As a representative reactor neutrino experiment we consider the operating parameters of the Daya Bay far detectors, which operate at 1~km baseline with a flux peaking at around 5~MeV~\cite{bay2007precision}. The neutrino wave packet width expected in beta decay was studied in detail in Ref.~\cite{jones2023width}.  The emitted neutrinos each have a different expected wave packet width that depends on the decaying isotope and the kinematics of the entangled final state particles, with $\sigma_{\nu,x}\sim10-400$~pm widths expected. The Daya Bay collaboration has published an experimental lower limit on the effective flux-averaged neutrino wave packet width~\cite{daya2017study}, though it does not exclude the currently predicted value.  Fig.~\ref{fig:Experiments} shows the red lines indicating the baselines of each of the reactor experiments described in~\cite{mention2011reactor} plus Daya Bay~\cite{an2012observation}, RENO~\cite{choi2020search} and Double Chooz~\cite{abe2012reactor} experiments, alongside the approximate energy spread of the reactor anti-neutrino spectra.

\begin{figure}[t]
\begin{centering}
\includegraphics[width=0.99\columnwidth]{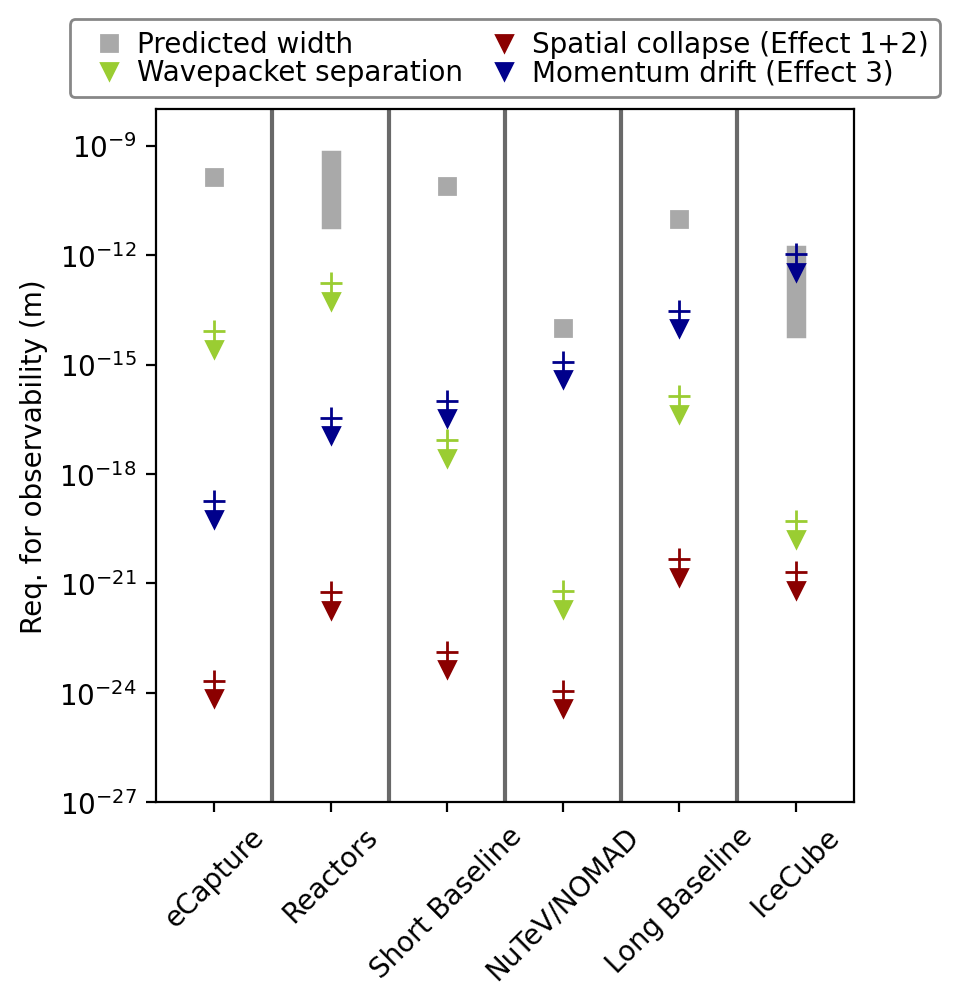}
\par\end{centering}
\caption{Wave packet size required for observability of the effects in each experiment. The wave packet must be smaller than the indicated value for the effect to be observably large. The threshold for observability is conservatively considered to be the point when the decoherence distance is ten times the experiment baseline.  \label{fig:Observability}}
\end{figure}

    \item {\bf Short Baseline Experiments:} Short baseline neutrino oscillation experiments such as MiniBooNE~\cite{aguilar2018significant}, MicroBooNE~\cite{abratenko2022search}, SBND~\cite{tufanli2017sbnd} and ICARUS~\cite{tortorici2019icarus}~ (shown as orange lines in Fig.~\ref{fig:Experiments}) detect neutrinos produced in beams of magnetically focused hadrons created by proton collisions on solid targets.  The charged hadrons travel through air as they decay, and interactions with the air molecules generate entanglements that quantum mechanically localize the neutrino parent. The emerging neutrino wave packet width has been calculated in Ref.~\cite{jones2015dynamical}.  For illustration of the scale of effect at these experiments we use the MiniBooNE experiment, which sits at a baseline of 500~m from the Fermilab Booster Neutrino Beam~\cite{acciarri2015proposal}, with a flux-averaged detected neutrino energy of around 500~MeV~\cite{aguilar2009neutrino}.  The expected  wave packet width for neutrinos of this energy in a conventional neutrino beam is approximately $8\times10^{-11}$ m~\cite{jones2015dynamical}.

    \item {\bf NuTeV/NoMAD:} Historically, much higher energy short baseline neutrino experiments have been operated using conventional neutrino beams. The NuTeV experiment~\cite{avvakumov2002search} operated with a neutrino flux spanning a range of 20-180~GeV at a baseline of 1420~m at Fermilab. NOMAD operated with a similar beam energy range  with an 825~m  baseline at CERN. Both are shown as purple lines on Fig.~\ref{fig:Experiments}.  We take 100~GeV as a representative neutrino energy in these experiments for subsequent calculations. The beam production process resembles that of the short baseline experiments, albeit at much higher energies. The results of Ref.~\cite{jones2015dynamical} imply a wave packet width for these experiments of approximately  $8\times10^{-14}$ .
    
    \item {\bf Long Baseline Experiments:} Accelerator neutrino experiments operating at longer baselines have a natural advantage when searching for weak, distance-dependent decoherence processes.  Examples include MINOS~\cite{evans2013minos}, OPERA~\cite{acquafredda2009opera}, NOvA~\cite{acero2018new}, T2K~\cite{abe2011t2k} and DUNE~\cite{abi2020long}, with baselines and approximate neutrino energy ranges shown as green lines on Fig.~\ref{fig:Experiments}.  The neutrino fluxes for these experiments are produced using the conventional hadron beam method, so the expected wave packet widths follow expectations from Ref.~\cite{jones2015dynamical}.  As a representative example we consider a 5~GeV beam propagating 735~km, reflecting the approximate operating parameters of the MINOS experiment. In these conditions the expected wave packet width is approximately 10$^{-11}$~m.  
    
    \item {\bf IceCube:} The longest baseline and highest energy neutrino oscillation analyses currently available come from IceCube, which recently set strong limits on anomalous decoherence in Ref.~\cite{icecube2024search}.  The IceCube neutrino baselines depend on zenith angle and span a range of values from a few hundred km to the diameter of the Earth, 12,000~km.  The energy spectrum of the sample studied in Ref.~\cite{icecube2024search} peaks at 1~TeV, and the energy span of the IceCube samples is roughly indicated as a shaded blue box in Fig.~\ref{fig:Experiments}.  The neutrino wave packet width expected from production in atmospheric air showers has not yet been calculated explicitly.  Since the particles are produced in pion and kaon decay, a rough scale can be inferred from the accelerator beamline calculations of~\cite{jones2015dynamical}, though with large error bars due to the rather different density of the atmosphere to that of the accelerator beam pipes. On the basis of those estimates we consider a viable range of plausible wave packet widths to be between $10^{-14}~\mathrm{m}<\sigma_{\nu,x}<10^{-12}~\mathrm{m}$.

\end{itemize}

The maximal wave packet widths for observability of natural Penrose-model decoherence effects are shown in Fig.~\ref{fig:Observability}, compared against the calculated or estimated wave packet widths in the relevant experiments. No effect is expected from the mass curvature effect (\textbf{effect~1}), which is in all cases too low to show on the figure. Neither is any experiment sensitive to pure position-space collapse 
 (\textbf{effect~2}).  However, the momentum drift (\textbf{effect~3}) that is analogous to the expected heating effect in germanium experiments~\cite{donadi2021underground,arnquist2022search} appears to within the IceCube sample sensitivity.  We also show for comparison the wave packet width required for the standard wave packet separation effect encoded in Eq.~\ref{eq:CohLength}.  For the lower energy experiments such as electron capture and reactors, we find that wave packets will already be far separated by the time the Penrose-like collapse process could occur, so it is intrinsically unobservervable in oscillations of these neutrinos. Due to the different energy and baseline scaling of the gravitational collapse vs wavepacket separation effects, for higher energy experiments the gravitational collapse becomes dominant over the standard wave packet separation effects.

\begin{figure}[t]
\begin{centering}
\includegraphics[width=0.99\columnwidth]{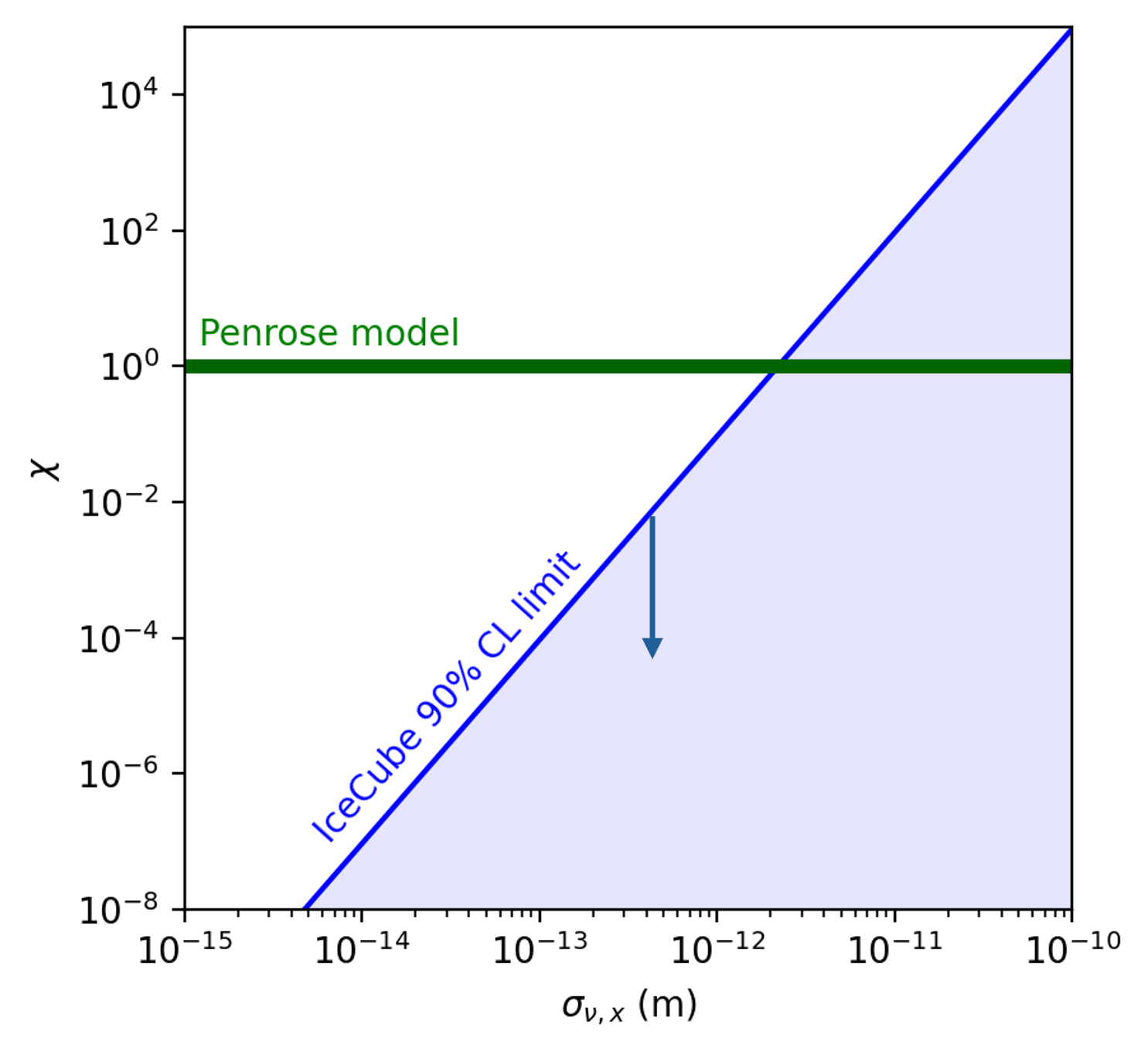}
\par\end{centering}
\caption{Limit on the effective decoherence scaling parameter $\chi$ compared against the Penrose model expectation $\chi\sim 1$. The IceCube limit on $\Gamma$ from Ref.~\cite{icecube2024search} has been re-expressed as a constraint on $\chi$ and $\sigma_\nu,x$ by noting that this is a decoherence mechanism scaling as $\Gamma\propto E^{1/2}$, at which the coherence distance at 1~TeV has been constrained to be less than $L_{coh}\leq 30$ Earth diameters. \label{fig:LimitPlot}}
\end{figure} 

To consider the quantitative extent to which this model is addressed by IceCube data, we note that the neutrino energy dependence of \textbf{effect~3} is that the coherence distance scales as 
\begin{equation}
L\sim4\sqrt{\frac{\sigma_{\nu,x}^{3}}{GE}},
\end{equation}
which corresponds to a model with $\Gamma$ scaling as $E_\nu^{1/2}$ in the notation of Ref.~\cite{icecube2024search}. The 90\%  confidence level (CL) limit on the coherence length $L_{90}$ under such a model can be evaluated based on the information provided in Ref.~\cite{icecube2024search} to be approximately 30 Earth diameters at a neutrino energy of $E_\nu\sim1$~TeV. To quantify the strength of the limit on the Penrose model, a scaling parameter $\chi$ is inserted into Eq~\ref{eq:PenroseUncert} to represent the strength of decoherence relative to the Penrose prescription, with $\chi\sim1$ corresponding to the natural Penrose model and $\chi>1$ representing stronger collapse effects, as 
\begin{equation}
\tau\sim\frac{ 1 }{\chi E_{g}}.
\end{equation}

\begin{figure}[t]
\begin{centering}
\includegraphics[width=0.99\columnwidth]{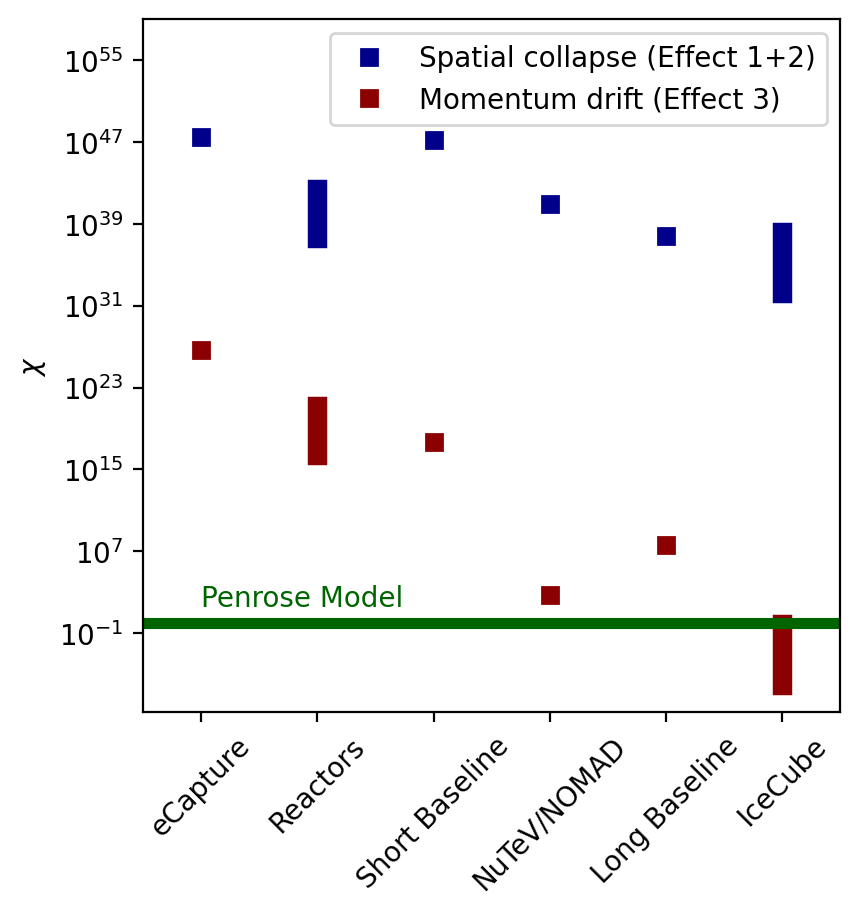}
\par\end{centering}
\caption{Effective constraint on the Penrose model strength parameter $\chi$ for cases where the coherence length is constrained to be 10 times the size of the experimental baseline. \label{fig:PenroseStrength}}
\end{figure}

The limit obtained in Ref.~\cite{icecube2024search} can then be re-expressed as an excluded region in the space of $\sigma_{\nu,x}$ and $\chi$, via 
\begin{equation}
\chi\leq\frac{16\sigma_{\nu,x}^{3}}{GE_\nu L_{90}^{2}}.
\end{equation}
The allowed region is shown in Fig.~\ref{fig:LimitPlot}. The neutrino wave packet would need to be larger than 2$\times 10^{-12}$~m for the spontaneous collapse process to be unobservable in IceCube. To conclusively rule out this possibility, a full calculation of the expected neutrino wave packet width in atmospheric neutrino oscillation experiments would be required, though based on past calculations a value wider than 2$\times 10^{-12}$~m seems unlikely. We defer a rigorous calculation of the wave packet width in atmospheric neutrino production processes to future work.

The expected effective strength $\chi$ that can be accessed in various experiments through both the spatial (\textbf{effect~2}) and momentum (\textbf{effect~3}) mechanisms are shown in Fig.~\ref{fig:PenroseStrength}, evaluated for hypothetical cases where coherence length limits are conservatively set to 10 times the oscillation baseline.  It is striking that the $y$ axis on this plot spans more than sixty orders of magnitude. The momentum diffusion effect is much larger than the position collapse effect, since the former does not require the wave packets to separate substantially before it has an impact - the fact it restricts the spatial extent of the superposition necessary imposes momentum uncertainty, via the Heisenberg principle.
Once again we observe that the higher energy experiments have a clear advantage, with IceCube spanning sensitivities beyond the Penrose natural model for the momentum-space effect.  The NuTeV/NOMAD class of experiments come close this benchmark as well, approaching the sensitivity of the far longer baseline IceCube search due to their smaller expected wave packet widths.   We note that a re-analysis of the data from those experiments in the context of the Penrose model could exceed the performance estimate presented here, due to the significant simplifying assumptions made in our estimates.

\section{Conclusions~\label{sec:Conclusions}}

We have studied the expected effects of the Penrose gravitational collapse model on neutrino oscillations. While superficially similar to the continuous localization models commonly employed in neutrino decoherence analyses, the Penrose model introduces additional subtleties into the calculation by coupling the geometrical form and separation of the neutrino wave packet to its collapse rate. 

We identify two main contributions to the expected decoherence rates under this model. The first is a purely spatial collapse which is similar to the mechanism originally proposed to forbid objects in macroscopic superpositions. This effect is dominated by the contribution from separation of the wave packets in space rather than the difference in their masses, but is very far from being observable given expected wave packet sizes in realistic neutrino emission processes. The second is a diffusive effect on the central neutrino momentum which leads to loss of oscillation phase coherence as the neutrinos are localized spatially.  This effect is unobservable in the majority of experiments, but is addressed by IceCube decoherence analysis~\cite{icecube2024search} at 90\% CL as long as the wave packet size at production in air showers is less than $\sigma_{\nu,x}\leq2\times10^{-12}$~m.

\section{Acknowledgements}

We are grateful to Stuart Hameroff and co-organizers for hosting The 2024 Science of Consciousness conference, where this paper was conceived and written.  We thank to Maria Startseva and Tyler Bryan for a series of beneficial Transmissions, and Tom Stuttard, Carlos Arg\"uelles, Jonathan Asaadi and Varghese Chirayath for useful comments on this manuscript.  BJPJ is supported by the US Department of Energy under awards DE-SC0019054 and DE-SC0019223, and OS under award DE-SC0022296.

\bibliographystyle{unsrt}
\bibliography{biblio}

\end{document}